\documentclass[twocolumn,showpacs,preprintnumbers,amsmath,amssymb]{revtex4}
\usepackage{graphicx}
\usepackage{dcolumn} 
\usepackage{bm}      
\usepackage{epsf,psfig}
\newcommand{\la}{\langle}
\newcommand{\ra}{\rangle}
\newcommand{\ua}{\uparrow}
\newcommand{\da}{\downarrow}

\newcommand{\om}{\omega}

\begin{document}


\title{Comment on ``Gapless Spin-1 Neutral Collective Mode Branch for Graphite''}

\author{N. M. R. Peres$^1$, M. A. N. Ara\'ujo$^2$ and A. H. Castro Neto$^3$}

\affiliation{$^1$Departamento de F\'{\i}sica 
e GCEP, Universidade do Minho, P-4710-057, Braga, Portugal,}
\affiliation{$^2$Departamento de F\'{\i}sica, 
Universidade de \'Evora, P-7000, \'Evora, Portugal,}
\affiliation{$^3$
Department of Physics, Boston University, 590 Commonwealth Ave.,
Boston, MA, 02215}
\maketitle

The interest in strongly correlated systems in frustrated lattices 
has been increased recently due to the possible realization of exotic
magnetic states \cite{anderson}, spin and charge separation in two 
dimensions \cite{matthew}, and the discovery of superconductivity
in Na$_x$CoO$_2$.$y$H$_2$O \cite{tanaka}. 
In a recent paper \cite{baskaran}, Baskaran and 
Jafari have proposed the existence of a neutral spin collective mode 
for a graphene sheet that is modeled as a half-filled Hubbard model
in the honeycomb lattice. These calculations were based on two main
approximations: the random phase approximation (RPA), and the 
{\it single cone approximation}. 
Since inelastic neutron scattering can be used to study this
spin collective mode in graphite, we decide to revisit the problem
without making use of the single cone approximation but considering
the entire band structure. We found that
such a spin collective does not in fact exist within RPA, being a
consequence of the single cone approximation.   

\begin{figure}[tbh]
\includegraphics[width=6cm,keepaspectratio]{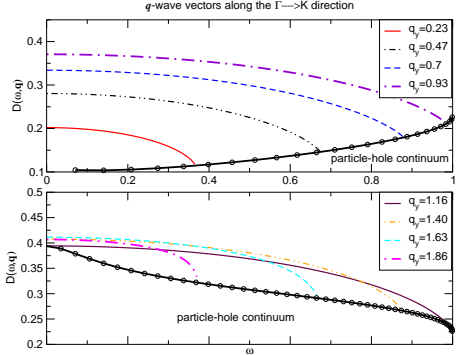}
\caption{Transverse susceptibility denominator $D(\omega,\bm q)$, 
Eq. (\ref{pole}), as function of $\omega$, for $U=2$, considering
eight  different  $\bm q$-vectors
along the $\Gamma$ to $K$ direction in the Brillouin Zone. 
For each wavevector, the line terminates
 at the point where $\omega$ enters the particle-hole continuum.}
\label{wscan}
\end{figure}

We treat the problem as two inter-penetrating triangular 
sub-lattices ($a$ and $b$) with a Hamiltonian given by: 
$
H= \sum_{\bm k,\sigma}[\phi(\bm k)a^\dag_{\bm k,\sigma}b_{\bm k,\sigma}+
H.c.] + H_U\,,$
where ${\bf k}$ is the momentum, 
$\vert \phi(\bm k)\vert = \epsilon_+(\bm k)$, 
$\epsilon_\pm(\bm k)= 
\pm t\sqrt{1+4\cos(ak_x\sqrt 3 /2) \cos(ak_y/2)+4\cos^2(ak_y  /2)}$
is the electronic dispersion,
and $H_U = \sum_i U n_{i,\uparrow} n_{i,\downarrow}$. 
The transverse spin susceptibility is defined as: 
$
\chi_{+-}^{\alpha,\gamma}=
\int_0^\beta d\,\tau\sum_{\bm p,\bm k}e^{i\om_n\tau}
\la T_\tau \alpha^\dag_{\bm p+\bm q,\ua}(\tau)\alpha_{\bm p,\da}(\tau)
\gamma^\dag_{\bm k-\bm q,\da}\gamma_{\bm k,\ua}\ra\,,
$
with $\alpha,\gamma=a,b$. In RPA the existence of two sub-lattices 
leads to a tensor form for the susceptibility given by 
$\bm{\chi}_{+,-}=\bm D^{-1} \bm{\chi}^0_{+,-}$, with 
\begin{equation}
\bm D=\left(
\begin{array}{cc}
1- u\chi_{+-}^{a,a;0} & -u\chi_{+-}^{a,b;0} \\
 -u\chi_{+-}^{b,a;0} & 1-u\chi_{+-}^{a,a;0} 
\end{array}
\right)
\end{equation}
with $u=U/N$. The matrix elements $\chi_{+-}^{\alpha,\gamma;0}$ of  
$\bm{\chi}^0_{+,-}$ are the non-interacting
susceptibilities. For $n=1$ and $T=0$ we have $
\chi^{a,a;0}_{+,-}=\chi^{b,b;0}_{+,-}=-\frac 1 2\sum_{\bm k}
C(\bm k,\bm q)$ and $
\chi^{a,b;0}_{+,-}=[\chi^{b,a;0}_{+,-}]^\ast=\frac 1 2\sum_{\bm k}
F(\bm k,\bm q)
C(\bm k,\bm q)\,,
$ with $F(\bm k,\bm q)=
[\phi^\ast(\bm k)\phi(\bm k-\bm q)]/[
\vert \phi(\bm k)\vert\vert\phi(\bm k-\bm q)\vert]$ and 
$C(\bm k,\bm q)=[\epsilon_+(\bm k)+\epsilon_+(\bm k-\bm q)]/[(i\omega)^2-
(\epsilon_+(\bm k)+\epsilon_+(\bm k-\bm q))^2]$.
The collective magnetic modes are determined from $\det[\bm D] =0$, or
\begin{equation}
D(\omega,\bm q)=1-2u \chi_{+-}^{a,a;0}+u^2 
(\chi_{+-}^{a,a;0})^2-u^2\chi_{+-}^{a,b;0}\chi_{+-}^{b,a;0}=0.
\label{pole}
\end{equation}
Collective modes are only well defined outside the particle-hole continuum 
otherwise these modes become Landau damped.  We searched for well defined 
magnetic modes, $\omega(\bm q)$,  below the continuum
of particle-hole excitations, and found no solutions for any value of $U$. 
In  Fig. \ref{wscan} we plot $D(\omega,\bm q)$ for eight different 
$\bm q$-vectors and $\omega$ ranging from zero to the point where the
particle-hole continuum begins.  As a consequence, the full structure 
of the Hubbard model's RPA susceptibility in the honeycomb
lattice does not show a collective magnetic mode as proposed in ref.
\cite{baskaran}.
We have also made two other independent checks of Eq. (\ref{pole}): 
(i) in the limit $\bm q,\omega\rightarrow 0$ we obtain the 
well known Hartree-Fock critical value
$U_c$ for the antiferromagnetic instability \cite{sorella}; (ii) 
when used in the antiferromagnetic ground state,  Eq. (\ref{pole}) 
does give the correct spin-wave spectrum \cite{peres}.

\end{document}